\documentclass[iop,numberedappendix] {emulateapj}

\usepackage{amsmath}
\usepackage{mathrsfs}
\usepackage{empheq}
\usepackage{url}
\usepackage{multirow}
\usepackage{dcolumn}   
\usepackage{amssymb}   
\usepackage{hyperref}
\usepackage[applemac]{inputenc}

\usepackage{graphicx}
\usepackage{bm}
\usepackage{color}

\shorttitle{the Hall effect}
\shortauthors{Fulvia Pucci, Marco Velli, Anna Tenerani}

\begin{document}


\title{Fast Magnetic Reconnection: ``Ideal" Tearing and the Hall Effect}
\author{Fulvia Pucci\altaffilmark{1,2},  Marco Velli\altaffilmark{3}, Anna Tenerani\altaffilmark{3}}
\email{fpucci@roma2.infn.it}
\email{fulvia.pucci@nifs.ac.jp}
\email{mvelli@ucla.edu}
\email{annatenerani@epss.ucla.edu}
\affil{\altaffilmark{1} Dipartimento di Fisica e Astronomia, Universit\`a di Roma Tor Vergata}
\affil{\altaffilmark{2} National Institute for Fusion Science, National Institutes of Natural Sciences,
Toki 509-5292, Japan}
\affil{\altaffilmark{3} Earth, Planetary and Space Sciences, University of California, Los Angeles}

\begin{abstract}
One of the main questions in magnetic reconnection is the origin of triggering behavior with on/off properties that accounts, once it is activated, for the fast magnetic energy conversion to kinetic and thermal energies at the heart of explosive events in astrophysical and laboratory plasmas. Over the past decade progress has been made on the initiation of fast reconnection via the plasmoid instability and what has been called ``ideal" tearing, which sets in once current sheets thin to a critical inverse aspect ratio $(a/L)_c$: as shown by Pucci and Velli (2014), at $(a/L)_c \sim S^{-1/3}$ the time scale for the instability to develop becomes of the order of the Alfvén time and independent of the Lundquist number (here defined in terms of current sheet length $L$). However, given the large values of $S$ in natural plasmas, this transition might occur for thicknesses of the inner resistive singular layer which are comparable to the ion inertial length $d_i$. When this occurs, Hall currents produce a three-dimensional quadrupole structure of magnetic field, and the dispersive waves introduced by the Hall effect accelerate the instability.  Here we present a linear study showing how the "ideal" tearing mode critical aspect ratio is modified when Hall effects are taken into account, including more general scaling laws of the growth rates in terms of sheet inverse aspect ratio: the critical inverse aspect ratio is amended to $a/L \simeq (di/L)^ {0.29} (1/S)^{0.19}$, at which point the instability growth rate becomes Alfv\'enic and does not depend on either of the (small) parameters $d_i/L, 1/S$.  We discuss the implications of this generalized triggering aspect ratio for recently developed phase diagrams of magnetic reconnection.
\end{abstract}

\section{Introduction}

Magnetic reconnection is generally believed to be the mechanism responsible for explosive events in astrophysical, space and laboratory plasmas \citet{YamadaKulsrudJi:2010} \citet{GonzalesParker:2016}. Indeed this process allows the magnetic field to access new topologies, dissipating magnetic energy into heat and accelerated particles. However, in order for the energy to be initially stored in the field - a requirement for bursty, intermittent energy release to be possible - reconnection can not be occurring all the time: as measured with a clock based on the characteristic ideal dynamical time-scale, reconnection and resistive instabilities must have an off-on character.  A complete understanding of magnetic reconnection therefore requires explaining how energy accumulates in the magnetic field, how the configuration suddenly become unstable, and how magnetic energy release occurs on very fast time scales; this implies a study of dynamics over multiple length and time scales. Since natural plasmas are characterized by magnetic Reynolds numbers which can vary by many orders of magnitude, e.g. from $10^6$ in Tokamaks up to $10^{13}$ in the solar corona,  
the dissipation scales can often approach kinetic (ion-inertial or gyroradius) scales, so a lot of work has been devoted 
to the role of kinetic effects in explaining fast magnetic reconnection \citet{Hosseinpour_1}.
Recently it has been shown \citet{PucciVelli2014} that in high Reynolds number plasmas, and in particular for the ideal limit, $S\rightarrow \infty$, resistive instabilities survive and can become ideal, i.e. grow on the Alfv\'en time-scale, once an inverse aspect ratio $a/L= S^{-1/3}$ is reached. For such current sheets, which are thicker than the Sweet-Parker sheets, ($a/L= S^{-1/2}$), the maximum growth rate does not depend on S, which means that the trigger condition is, in some sense, "independent of the environment" (IT, for ``ideal" tearing). This was confirmed by numerical simulations \citet{Landietal2015},  \citet{Teneranietal2015a}. In particular in \citet{Teneranietal2015a}, where a collapsing current sheet at high Reynolds numbers was simulated, it was shown that, once the critical aspect ratio is reached, the instability takes place on ideal timescales developing multiple plasmoids, from which a hierarchy of secondary IT instabilities takes place.

Even if the inverse aspect ratio of a collapsing current sheet never reaches the Sweet-Parker thickness, kinetic effects may become important at large enough Lundquist numbers if the typical collisional resistive scales are smaller than either the ion inertial length or the thermal ion gyroradius \citet{CassakDrakeShay:2006, MalakitCassakShayDrake:2009}. When characteristic scales approach the ion inertial length, the Hall effect starts to play a role, and a quadrupolar magnetic field emerges, affecting magnetic reconnection dynamics. Indeed there is now copious evidence that Hall reconnection occurs, both from magnetospheric observations \citet{Mozeretal2002,Wygantetal2005, HallEastwood, Franketal:2016}, at the dayside magnetopause \citet{Vaivadsetal2004} and in the near-Earth magnetotail \citet{Runovetal2003, Borgetal2005,Nakamuraetal2006}, as well as in laboratory experiments \citet{Cothran2005, Ren2005,Yamada2006, Tharpetal:2013, Kaminouetal:2016} . 

It is important to recognize that the Hall effect intervenes well before the overall current sheet thickness approaches $d_i/L$, because the Hall term will affect the dynamics already when the inner, tearing mode resistive singular layer thickness approaches ion scales. As we shall see in the following sections, this leads to results which remain consistent with previous works, by reason of the properties of ``ideal'' tearing. In this paper we will consider the case of a planar configuration without a guide field:
the effect of a finite ion inertial length in the framework of the tearing instability was first carried out by \citet{Terasawa}, where a schematic illustration of the effect of a quadrupole magnetic field on the planar configuration was shown.

Terasawa (1983) demonstrated that the Hall effect produced a growth rate enhancement at high Lundquist numbers, but the time for the instability to develop was still very slow because the current sheet thickness was assumed to be macroscopic. On the other hand, when the ion inertial length $d_i$ becomes of the order of the length scale which characterizes the equilibrium magnetic field gradient $a$, ions and electrons decouple and whistler waves form, so two-fluid effects should be taken into account, and the growth rate should be scaled using typical whistler frequencies. 
Starting from these considerations, the regime where the inner resistive layer thickness becomes of the same order of the ion inertial length is investigated here, in sheets where the inverse aspect ratio scales as powers both of the inverse Lundquist number $a/L= S^{-\alpha}$ and of the normalized ion inertial length $d_i/L$. The idea is to recover a specific scaling for which the growth rate becomes independent both of the inverse Lundquist number and the ion inertial length, i.e. intrinsically fast, independently of the dominant small parameter driving the instability. 

The plan of the paper is the following: in section two we introduce the Hall effect and summarize the scaling relations of the fastest growing modes for current sheets with fixed aspect ratios. Section 3 then generalizes the ``ideal" tearing instability criterion, discussing how the corresponding scaling relations are affected by a finite ion inertial length, and confirming them with numerical solutions of the eigenvalue equations. In the conclusions we discuss this result and place them into context.  
\section{\label{sec:II}The tearing instability in the presence of the Hall effect.}
In this paragraph we summarize the classical stability problem of an equilibrium magnetic field configuration aligned along the $x$ axis (\,$\hat i$\,) and dependent on the perpendicular coordinate $y$ (\,$\hat j$\,)  in the form of a Harris current sheet,
\begin{equation}
\label{harris}
\vec B(y) = B(y)\hat{i}= B_0\, F (y/a)\hat{i}
\end{equation}
where $F \left( y/a \right)= tanh \left( y/a \right)$. If there is no out of plane (guide) field, a corresponding equilibrium pressure profile $p(y) =p_0 - B^2(y)/ 8\pi$ is required to guarantee equilibrium. Quantities are assumed to be uniform in the direction perpendicular to the plane of the equilibrium magnetic field, i.e. 
$\partial /\partial z=0$ but the perturbed fields are 3D, so the reconnection region may develop a spatial 3D structure. That the velocity and magnetic field perturbation along $y$ and $z$ directions couple, may be seen by writing out the generalized Ohm's law including the Hall term, \small
\begin{eqnarray}
\label{GenOhmslaw}
&{\vec E}+\dfrac{1}{c} ({\vec v} \times {\vec B})=\dfrac{4 \pi}{c^2} \eta_m {\vec J} +
\dfrac{1}{enc}\left({\vec J} \times {\vec B} \right),
\end{eqnarray}
\\[- 1.2ex]
\normalsize
where  $\vec E$ is the electric field, $\vec v$ the velocity, $\vec J$ the current, $\eta_m$ the magnetic diffusivity and $n$ the electron and ion density. In this form, Ohm's law includes the Hall effect and collisional resistivity but neglects electron pressure and electron inertia: the effect of these terms on``ideal" reconnection regimes has been discussed in \citet{DelSartoetal2016}. 

Upon linearization, the equations are non-dimensionalized using the magnetic field intensity $B_0$ (the perturbed magnetic field $\vec b = \vec b/B_0$), the magnetic field gradient scale $a$ (essentially the Harris sheet thickness), and the Alfv\'en time $\tau_A= a/v_A$ where the Alfv\'en speed $v_A = B_0/\sqrt{4\pi\rho}$.  Wavenumbers  $k$ along $x$ are also scaled with $a$ and we introduce the non-dimensional displacement $\xi_y= iv_y/(\gamma a)$. The growth rate is normalized to  $\tau_a$; consequently, the Lundquist number is defined as $S= a v_A/ \eta$. Denoting derivatives with respect to the non-dimensional coordinate ${y}=y/a$ with $'$, the tearing mode equations become:
\small
\begin{eqnarray}
\footnotesize
&& (\xi_y''-k^{2}\xi_y)=-{k}\dfrac{1}{ \tilde{\gamma}^2} \, \left[F({b}_y''-{k}^{2}{b}_y)-F''{b}_y\right]\label{classichall1}\\[2ex]
&&{\xi}_{z} =  - \dfrac{1}{ \tilde{\gamma}^2} \,  F  {k} {b}_{z}\label{classichall2}\\[2ex]
&&{ b}_y = {k}\, F {\xi}_y +  \dfrac{1}{S\tilde{\gamma}} ({b}_y'' -{k}^2 {b}_y)- \dfrac{h}{\tilde{\gamma}} \,F{k}^2 {b}_z\label{classichall3}\\[2ex]
&&{ b}_z= {k}\, F {\xi}_z +  \dfrac{1}{S\tilde{\gamma}}({b}_z'' -{k}^2  {b}_z)\nonumber\\
&&  \ \ \ \ \ \ \ \ \ \ \ \ \ \ \ \ \ \ \ \  -\dfrac{h}{\tilde{\gamma}} \left[F( {b}''_y  -{k}^2  {b}_y )- {b}_y F''\right].
 \label{classichall4}
\end{eqnarray}
\normalsize

Here  $h=d_i/a$ identifies the \textit{Hall coefficient} which couples the $z$ and $y$ components of the perturbed magnetic field, and when $h=0$ the classic tearing mode equations (FKR) are recovered. As mentioned above, we expect the Hall term to become important when $d_i$  becomes of the same order as the thickness of the internal, singular layer describing the resistive tearing mode. Since for the classic resistive tearing mode such thickness $\delta$ (also normalized to the shear-length $a$) , for the fastest growing mode, scales as $\delta \sim  S^{-1/4}$, we define 

\begin{equation}
{P}_h \equiv \dfrac{h}{\delta} \sim {h} {S}^{1/4},
\end{equation}
so that ${P}_h \sim 1$ means the Hall effect is no longer negligible. 

The system of equations (\ref{classichall1}-\ref{classichall4}) is a sixth order two-point eigenvalue problem (for given $k, h, S$) in which the solutions develop large gradients in $y$  around the $x$-axis with increasing $h, \, S$.  As mentioned above, the problem was first studied by \citet{Terasawa}, who showed that the tearing mode develops three characteristic layers: in addition to the sheet thickness $a$ and to the inner, singular layer, familiar from the purely resistive tearing mode \citet{FKR}, an intermediate layer arises, in which the Hall current effect is also essential. This intermediate layer complicates the asymptotic analysis of the problem, introducing a dependence of the parameter $\Delta'$, defined below, on $h$: it is therefore best to resolve the eigenvalue problem numerically using the Lentini-Pereyra method, \citet{lentini}, and compare the results to analytic estimations derived from an heuristic generalization of the classic tearing asymptotic matching (in the vein of \citet{Terasawa}). 

\begin{figure}
\includegraphics[width=90mm]{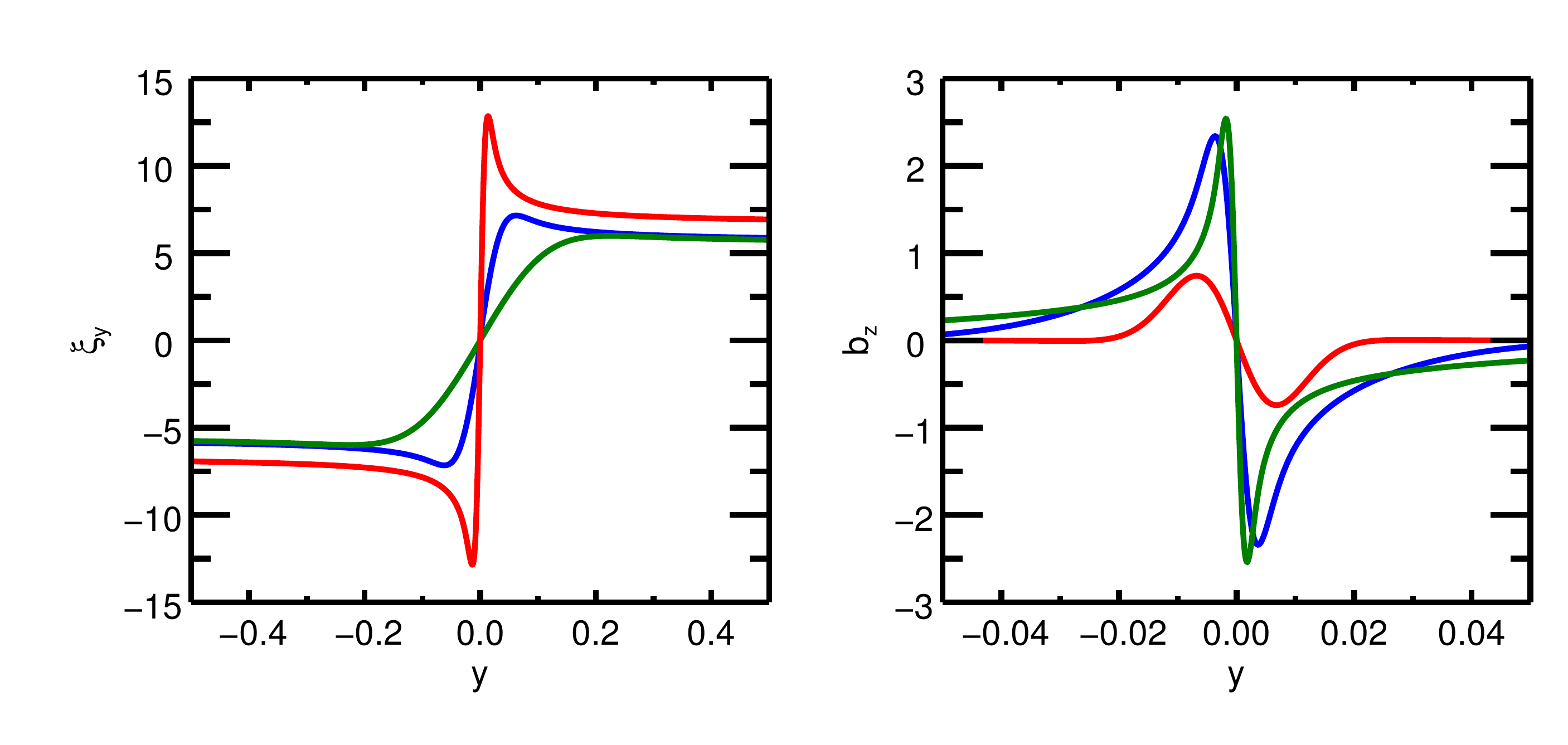}
\caption{Eigenfunction $\xi_y$ (left panel) and $b_z$ (right panel) for $P_h=1, 10, 40$, with $S=10^8$. Notice the different behavior of  $\xi_y$, whose peaks separate with increasing $P_h$ (i.e. ion inertial length since $S$ is fixed), and $b_z$, whose peaks converge.}
\label{xiybz}
\end{figure}

We first illustrate the two-layer development with finite $h$ by examining the behavior of the eigenfunctions $\xi_y$ and $b_z$. Recalling that $F$ is an odd function of $y$, it is easy to see that the ideal, marginal form of eqs. (\ref{classichall1}-\ref{classichall4}) yields solutions for $b_y$, $\xi_y$, $b_z$, $\xi_z$ that are respectively even, odd and odd, even. Also, at great distances from the $x$-axis, solutions decay exponentially. As proxies for the intermediate, Hall layer, and the inner singular layer around the $x$-axis, we use the distance between two peaks of the displacement $\xi_y$ and of the Hall generated field $b_z$, plotted in Fig.(\ref{xiybz}), left and right panels. 
The left panel shows the eigenfunction $\xi_y$ for fixed Lundquist number $S =10^8$ and  values $P_h=1, 10, 40$, in red, blue, green respectively: one sees how the profile widens with increasing $h$. Plotted in the right panel is the eigenfunction for $b_z$, which displays a different behavior, with a peak to peak central  thickness  which decreases with increasing $P_h$. It is not that $\xi_y$ does not display signatures of the internal, singular layer, it is just that it is less apparent, coming as it does in the form of an abrupt change in its gradient, rather than the more visible maximum/minimum that $b_z$ displays.

\begin{figure}
\includegraphics[width=80mm]{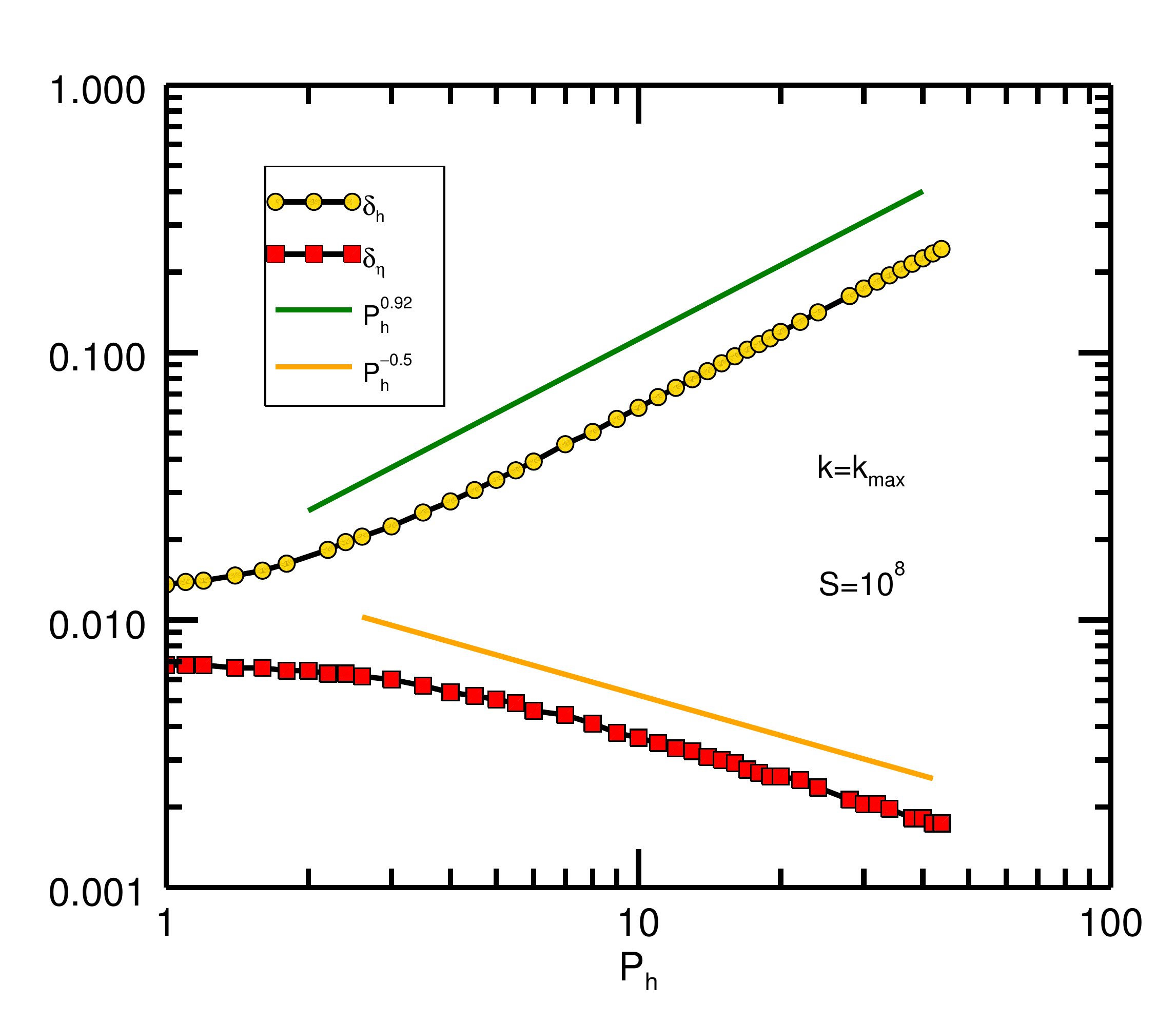}
\caption{Intermediate Hall layer and inner resistive layer as a function of the parameter $P_h$, for $S=10^8$. The intermediate thickness starts to increase as $P_h\ge1$, as shown in while the inner, resistive layer shrinks.}
\label{deltaofP}
\end{figure}

The thicknesses of both the intermediate layer and the inner resistive layer are plotted as a function of the parameter ${P}_h$ in Fig.\ref{deltaofP}. 
For very small $P_h$, one expects to recover the thickness of the resistive layer $\delta_\eta \sim S^{-1/4} \simeq 0.01$, while for increasing $P_h$ the intermediate layer thickness increases (almost linearly)  $\delta_h \sim P_h^{0.94} $ while the  resistive, singular layer is found to decrease with $P_h$ as $ \delta_\eta \sim P_h^{-1/2}$. The two curves in the figure do not appear to intersect at small $P_h$ because of the logarithmic scale (convergence would be at $-\infty$ on this scale).

To understand the thinning of the inner resistive layer let's generalize the classical tearing mode asymptotic matching theory heuristically,  taking into account the effects of $h$ ($P_h$).  In the classic, resistive tearing mode without Hall effect, the maximum growth rate may be obtained by matching the scalings obtained in the so-called large $\Delta'$ and small
$\Delta'$ regimes. $\Delta'$ is defined by searching for the solution of the linearized momentum equation neglecting
the growth rate, i.e. assuming marginal stability:
$$F({b}_y''-{k}^{2}{b}_y)-F''{b}_y=0. $$ Because the tearing mode is centered where the equilibrium magnetic field changes sign,
the correct solution vanishes at large $y$. It is easy to see then that the solution has a discontinuity in derivative as $y \rightarrow 0$, 
and
$$\Delta' = {\small \rm lim}_{y \rightarrow 0} {b'_y(y)-b'_y(-y)\over b_y(0)},$$ which for the particular equilibrium at hand gives $\Delta' = 2 (1/k -k)$.  The two regimes of the tearing mode are determined  by whether the product $\Delta' \delta>>1$ (the large $\Delta'$ regime, also known as the resistive kink regime) or $\Delta' \delta<<1$ (small $\Delta'$ or the classical tearing mode regime): depending on which of the two regimes obtains, the second derivative of $b_y$ within the inner singular layer scales either as ${b}_y'' \sim b_y /\delta^2$ (large $\Delta'$) or ${b}_y'' \sim b_y \Delta'/\delta$ (small $\Delta'$). As a result, the relationships between $\xi_y, b_y$ obtained by matching the inner to outer layer together with 
$\delta, \gamma$ may be summarized as follows:
\begin{eqnarray}
\nonumber
&{\rm  \small large}\, \, \Delta'\qquad& {\rm \small small}\, \, \Delta'\\ 
&b_y\sim k \delta \xi_y\qquad&  b_y\sim k \delta \xi_y \label{deltaprime1}\\
&b_y\sim b_y/(S\gamma\delta^2)\qquad& b_y\sim \Delta' b_y/(S\gamma\delta) \label{deltaprime2} \\
& \gamma^2\xi_y/\delta^2\sim kb_y/\delta\qquad & \gamma^2\xi_y/\delta^2 \sim k \Delta' b_y.
\label{deltaprime3}
\end{eqnarray}
Eq.(\ref{deltaprime1}) comes from the ideal terms (outer region) of the induction equation, eq.(\ref{deltaprime2}) from the inner, resistive layer of the induction equation,
while eq.(\ref{deltaprime3}) comes from the momentum equation estimated within the inner, resistive layer. From these equations one immediately finds the growth rate scalings for the large and small $\Delta'$ regimes respectively,
$\gamma \sim k^{2/3} S^{-1/3}$, $ \gamma \sim {\Delta'}^{4/5} k^{2/5} S^{-3/5}$.
With increasing $S$ at any fixed $k$ modes transition from large to small $\Delta'$, and the growth rate of the fastest growing mode
at $k(S), S$ can be found by matching the two dispersion relations, taking into account that in the large $\Delta'$ regime one has $\Delta' \sim 1/k$: $k_m \sim S^{-1/4}$,$ \gamma \sim S^{-1/2}$.

\begin{figure}
 \includegraphics[width=85mm]{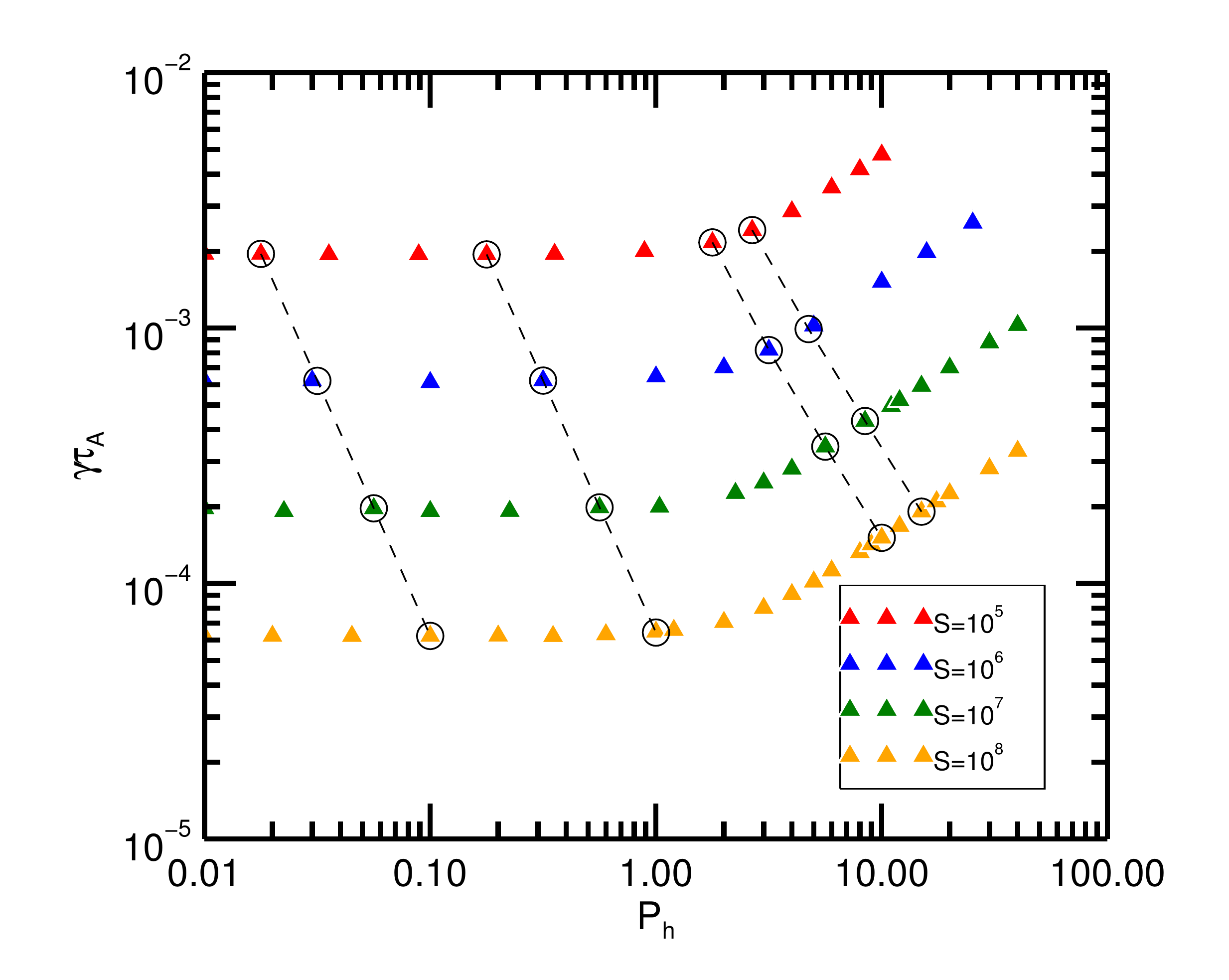}
\caption{Growth rate in Alfvén time units as a function of the parameter $P_h$, for $S=10^5,10^6,10^7,10^8$. The growth rate remains constant for $P_h<1$ and starts to increase as $P_h\ge1$, as expected.}
 \label{gammamaxh2}
\end{figure}

Extending this reasoning to include Hall terms requires analyzing the equation for the Hall field, though one must be careful 
in how to approximate $\xi_z$ and $b_z$ at the edge of the inner, resistive layer of thickness $\delta_\eta$. Because $b_z$ has the same symmetry as $\xi_y$,
the dimensional estimate for the second derivative is $b_z'' \sim -b_z/\delta^2$. A dimensional estimate from eqs.(\ref{classichall1}-\ref{classichall4}) then shows, following the strategy of eqs. (\ref{deltaprime1}-\ref{deltaprime3}) that the Hall term in the induction equation, i.e. the last term in eq.(\ref{classichall3}) has magnitude
\begin{eqnarray}
&{h^2k^2\over \gamma^2} {b_y \over 1+1/(S\gamma\delta_\eta^2) + k^2\delta_\eta^2/\gamma}, \, ({\rm large}\, \Delta') \nonumber \\
&{h^2k^2\delta\over \gamma^2} \Delta' {b_y \over 1+1/(S\gamma\delta_\eta^2) + k^2\delta_\eta^2/\gamma},  \,({\rm small}\, \Delta').
\label{hallbz}
\end{eqnarray} 
In the $b_y$ denominators above, the third term is always negligible, while the $S\gamma \delta_\eta^2$ term is either $<<1$, in the small $\Delta'$ regime, or O(1), in the large $\Delta'$ regime. In both cases, once  $P_h \geq 1$, 
the contribution of this Hall term contribution to  eq.(\ref{classichall3}) dominates compared to the ideal MHD (convective term, first one on the rhs) contribution, leading to dispersion relations of the form
\begin{eqnarray}
&\gamma \sim (hk) & \qquad \delta_\eta \sim (hkS)^{-1/2},\nonumber\\
&\gamma \sim (hk)^{1/2} \Delta' S^{-1/2}&\qquad \delta_\eta \sim (hkS)^{-1/2},
\label{halldeltap}
\end{eqnarray}
in the large and small $\Delta'$ regimes respectively. Again matching the two to obtain the scaling of $k$ for the maximum growth rate, taking into account that in the large $\Delta'$ regime one has $\Delta' \sim 1/k$,  one finds
\begin{equation}
k_m \sim h^{-1/3} S^{-1/3}\,, \, \gamma_m \sim h^{2/3} S^{-1/3}\,, \, \delta_\eta \sim h^{-1/3}S^{-1/3},
\nonumber
\label{maxhall}
\end{equation}
showing both that the growth rate of the instability is enhanced and that the resistive internal layer width decreases with increasing $h$.
The above scalings have been obtained neglecting the influence of $h$ on $\Delta'$, which, as already observed by Terasawa (1981) is too strong an approximation. Indeed the direct numerical resolution of the eigenvalue problem shows that the growth rate does not quite follow the derived scalings. To fit the numerical results, we use the approximation for maximum growth rate
\begin{equation}
\gamma_{max} {\tau}_A \sim \gamma_0 {S}^{-1/2} (1 + \gamma_1{P}^{\zeta}_h),
\label{Phscaling}
\end{equation}
where the well known scaling in the IT regime when Hall is negligible implies that 
$\gamma_0 \sim 0.62 $.
Numerical results for the maximum value of the growth rate as a function of $P_h$, with four different fixed values of the Lundquist number (i.e. variable $h$) are shown in Fig.\ref{gammamaxh2}.
Dashed lines join points with the same value of the Hall coefficient $h$: note that the dashed lines in the regimes of small $P_h$ and large $P_h$, while separately parallel, are not parallel across the transition at $P_h\simeq1$. This is because
the Hall term influences the scaling with $S$ via $P_h$ rather than simply $h$. Fitting the curve in Fig.(\ref{gammamaxh2}), for $P_h\gg1$ we obtain

\begin{eqnarray}
\label{growthratePhgrosso}
\gamma_{max} \tau_A = \gamma_{01} S^{-1/2} P^{\zeta}_h \nonumber \\
\gamma_{01} = 0.41 \pm 0.01, 
\end{eqnarray}
where $ \zeta = 0.564 \pm 0.007$ and $\gamma_{01} = \gamma_{0}\, \gamma_{1} $. A quick check with the heuristic estimate shows that while the exponent for $S$ is the same, $\zeta$ is significantly different, i.e. $\zeta \simeq 0.56$ rather than   $\zeta= 2/3\simeq 0.67$.

We want to also remark that for all of our calculations $h < 1$, i.e. $\dfrac{a}{d_i} > 1$, and  ${\tau}_W=\dfrac{a}{d_i} {\tau}_A > \tau_A$, which means energy conversion and outflow from the resistive region occur on Alfv\'enic timescales. The wavevector $k$ for which we have the maximum growth rate is not affected by $P_h$, i.e. even for $P_h\gg1$, $k_{max} (P_h) \sim constant$.\\
\section{\label{sec:II}Fast resistive magnetic reconnection in the presence of the Hall effect.}

\begin{figure} 
 \centering
\includegraphics[width=85mm]{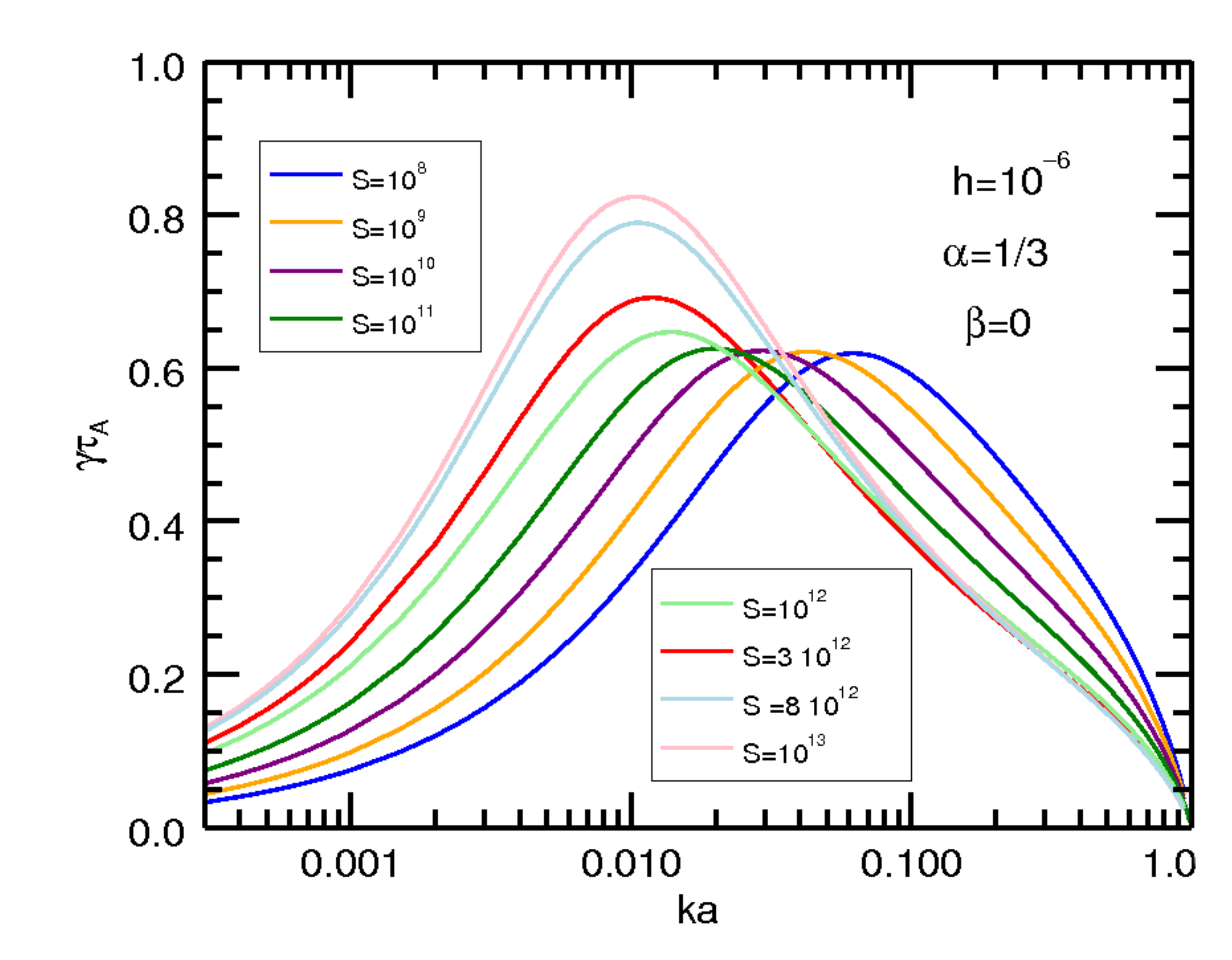}
\caption{Figure: Growth rate as a function of the wavevector $k$ for different Lundquist numbers for an inverse aspect ratio which scales as $a/L \sim S^{-1/3}$ and an Hall coefficient $h=10^{-6}$. The maximum growth rate is constant till the Hall effect becomes important i.e. $S=10^{12}$.}
\label{idealtearing}
\end{figure}

We come now to the question of how the critical resistive IT aspect ratio is modified
once the resistive layer thickness becomes comparable to the ion inertial length.  Following our previous work, we will
consider current sheets with a macroscopic length $L$, which will therefore be used to define the Lundquist number and Alfvén time: 
$S \equiv Lv_A/\eta$, $h \equiv d_i/L$. Let us first consider the resistive ``ideal" tearing mode at the critical current sheet aspect ratio,
scaling as $a/L \sim S^{-1/3}$. It was shown (Pucci and Velli, 2014) that in this case resistive inner layer scales as $\delta_\eta/ \sim S^{-1/2}$. It therefore seems logical to generalize the parameter defining the relevance of the Hall effect to
\begin{equation}
\label{Phbar}
P_h \equiv h S^{1/2}. 
\end{equation}
To confirm this, we first consider a sequence of equilibria at various aspect ratios at large $S$ and scaling as  $a/L \sim S^{-1/3}$, but fixing the value of $h=10^{-6}$. As $S$ increases, we expect the maximum growth rate to be constant until $ P_h \sim 1$, at which point the Hall effect acceleration should lead to an increase of the growth rate. This is because once $P_h \sim 1$ is passed, the scaling
$a/L \sim S^{-1/3}$ thins the current sheet too much, leading to the same paradox  of the plasmoid instability on SP sheets,\citet{loureiro07}, namely a growth rate which diverges with increasing $S$. This is confirmed by the numerical solution of the eigenvalue equations shown in Fig.\ref{idealtearing}, where at first (dark blue through dark green lines) one sees that the maximum growth rate is the same, independently of $S$, yet once $P_h >1$ is surpassed, the maximum growth rate rises again (light green through pink curves), rapidly increasing with increasing $S$. What this means is that once the Hall effect becomes important, the critical aspect ratio should no longer depend only on $S$, but also on the parameter $h$ or $P_h$, as there are now two asymptotic parameters at work.

We therefore generalize to an inverse aspect ratio which varies in the parameter space $(S,h)$, scaling as 
\begin{equation}
\dfrac{a}{L} \sim S^{-\alpha}P_h^{\beta}
\label{aspectratio}
\end{equation}
and search for exponents such that the time for the instability to develop becomes independent of the parameters of the system. Starting from Eq.(\ref{growthratePhgrosso}), we must now renormalize all quantities to $L$ rather than $a$:
\begin{eqnarray}
\gamma_{max} \tau_A \dfrac{a}{L}    \simeq    S^{-1/2}  \left( \dfrac{a}{L} \right)^{-1/2}  \left( S^{1/4} h \left( \dfrac{a}{L} \right)^{-3/4}  \right)^{\zeta}
\end{eqnarray}
where $\zeta$ was determined in the previous section numerically ($\zeta = 0.564 \pm 0.007$.
From the definition eq.(\ref{Phbar}), inserting the  explicit aspect ratio dependence from eq.(\ref{aspectratio})
we obtain
\begin{eqnarray}
&&\gamma_{max} \tau_A   \simeq S^{-1/2(1+\zeta/2)} P_h^{\zeta} \left( S^{-\alpha}P_h^{\beta} \right)^{-3/2(1+\zeta/2)} \nonumber\\
&&= S^{-1/2(1-3\alpha)(1+\zeta/2)} P_h^{\zeta-3/2\beta(1+\zeta/2)}.
\end{eqnarray}
We now search for the values of $\alpha,\beta$ that cancel the growth rate dependence on $S, P_h$. First, notice that the
growth rate is independent of the Lundquist number for $\alpha=1/3$, independently of the value of $\zeta$. Then, for $\beta$ we
may write:

\begin{eqnarray}
\beta = \dfrac{2}{3} \dfrac{\zeta}{(1+\zeta/2)},
\end{eqnarray}
which allows us to calculate the exponents for which $\gamma_{max} \tau_A \simeq 1$:
\begin{equation}
\alpha=1/3\,, \qquad \beta= 0.29, 
\label{exponentshall}
\end{equation}
and the error coming from the numerical determination of 
$\zeta$ is then  $\Delta \beta \simeq 3 \Delta \zeta = 0.02$. 
\section{\label{sec:IIB}Results.} 
\begin{figure}
\centering
 \includegraphics[width=90mm]
{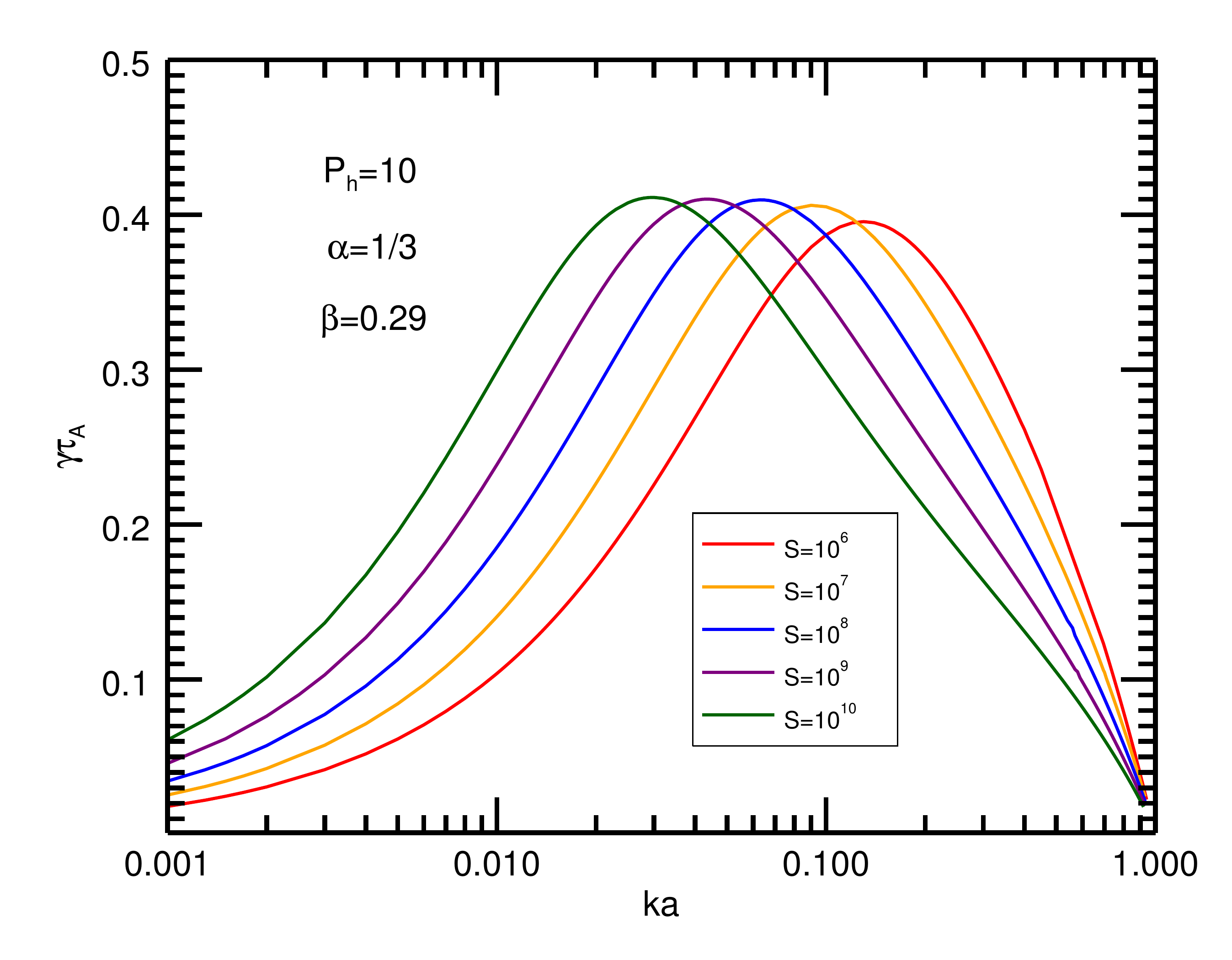}
\caption{Maximum growth rate in Alfvén time units as a function of the parameter $k$, for $P_h=10$ for different values of the Lundquist number. Notice that the maximum growth rate is constant.}
 \label{Phconstidelahall}
\end{figure}

\begin{figure}
 \includegraphics[width=90mm]{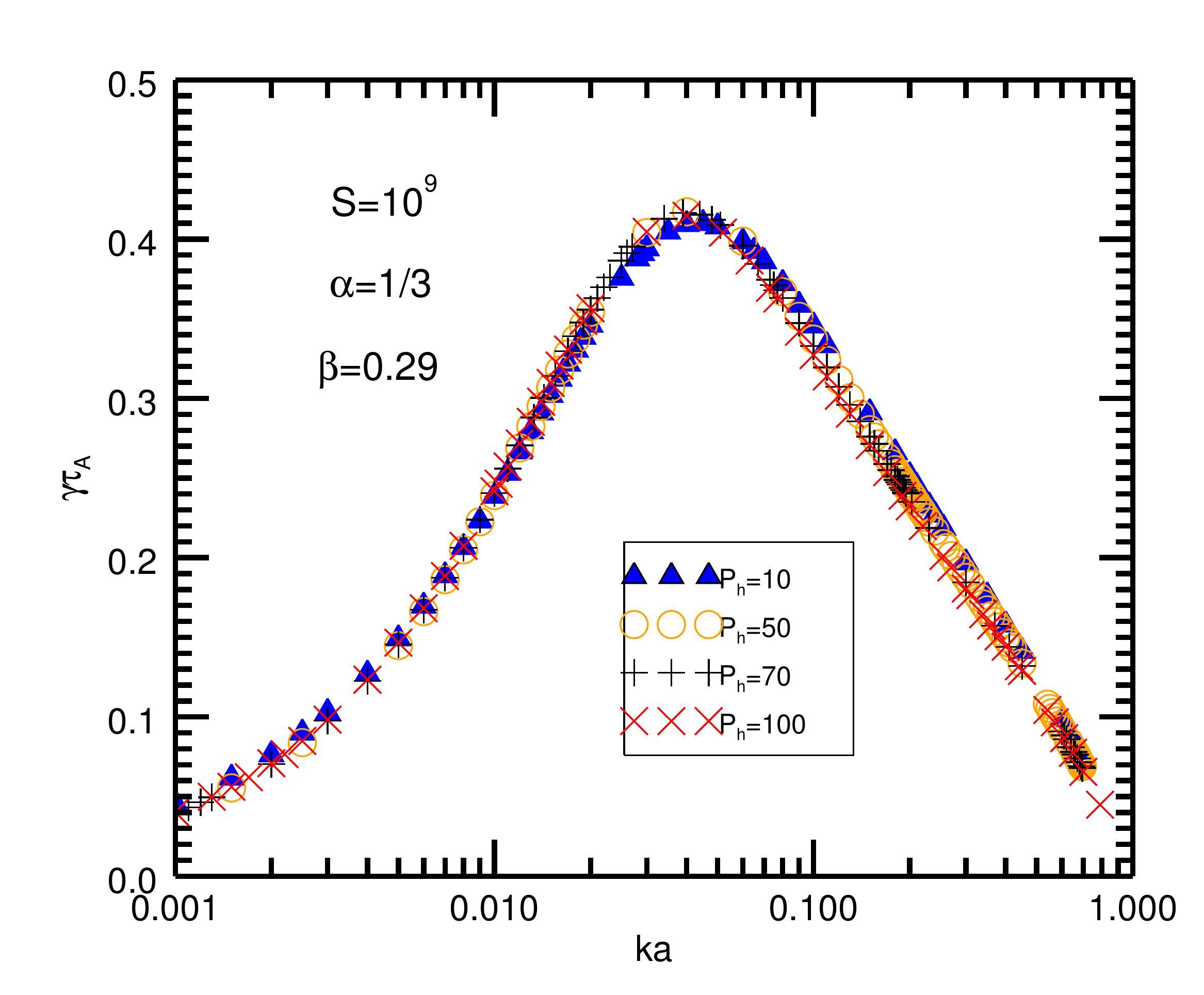}
\caption{Maximum growth rate in Alfvén time units as a function of the parameter $k$, for $S=10^9$ for different values of $P_h$. Notice that the maximum growth rate is constant.}
 \label{Sconstidelahall}
\end{figure}

The correctness of the prediction  for the scaling relation of the inverse aspect ratio eq.(\ref{exponentshall}) was verified numerically: 
first, in Fig.\ref{Phconstidelahall} we show solutions for the  growth rate as a function of wave-number for different values of $S$ and a  constant value $P_h \gg1$ ($P_h=10$). The asymptotic value of the growth rate is $\gamma_{max} \tau_A \sim 0.41=\gamma_{01}$, as expected for a growth rate independent of the parameters.
Second, we show the values of the growth rate, again as a function of $k$, at constant values of $S$ 
but different values of $P_h \gg 1$ (obtained by varying $h$), in Fig.\ref{Sconstidelahall}.  Notice how $\gamma_{max} \tau_A$ remains constant,$\gamma_{max} \tau_A =\gamma_{01}$, and there is no shift in wavenumber $k$ of the maximum growth rate with changes in $h$.
The fact that the coefficient $\gamma_{01} \simeq 0.41 < \gamma_0 \simeq 0.62$ does not have direct physical significance, stemming as it does from our definitions of aspect ratio scaling: as one may immediately verify, one could change the specific values of these coefficients by redefining the aspect ratio scaling by including an arbitrary multiplicative constant. What is important of course is the value of the scaling exponents $\alpha, \beta$ which determine the critical aspect ratios beyond which current sheets become so strongly unstable that they will never form.\\

Our choice for the normalization time,  $\tau_A$,  is valid only if $a/L \gg d_i/L$ so that $\tau_A < \tau_{\mathrm{w}}$ (the whistler timescale). We want to verify this hypothesis is satisfied by our critical aspect ratios. Starting from the Eq.(\ref{aspectratio}), we have
\begin{equation}
\dfrac{a}{L} \sim h^{\beta} S^{-\alpha+\beta/2} >h.
\end{equation}
which means that for $ S^{-1/2} \le h \le S^{-0.27}$ our hypothesis is verified, while for $h >S ^{-0.27}$ we should normalize the growth rate to the whistler time. These quantities have been listed in Tab. {\ref{tabella1}} for some astrophysical as well as for laboratory plasmas, where we assumed $\delta/L=S^{-1/2}$ and, if $P_h \ll1$, the inverse aspect ratio $a/L \simeq S^{-1/3}$, while if  $P_h \gg 1$, $a/L\simeq S^{-1/3} P_h^{1/3}$. 
\begin{table*}
\resizebox*{1.\textwidth}{!}{
\begin{tabular} {|c|c|c|c|c|c|}
\hline \hline
{\bf Plasma}                     &  {\bf Solar Cromosphere} & {\bf Solar Corona}         &{\bf Solar Wind}       &{\bf Magnetotail} 	                      &{\bf MRX} \\
\hline
{\bf n}                                      & $10^{14}$  		   &$10^{9}$  			& $3  $			&$0.1$                                               &$(2-6) \times 10^{13}$\\
{\bf L}                                       & $5\times10^{8}$ 	   & $2 \times 10^{9}$ 		& $10^{13}$               & $10^{9}-10^{10}$                   	& $6-20$\\
{\bf B}                                      & $50-200$  		   &$100$ 		                 &$10^{-4}$		&$10^{-4}$	                                &$(1-3) \times 10^{2}$\\
{\bf T}                                      &  $10^{4}$		           &$ 10^{6}$			&  $10^{5}$ 		&  $10^{7}$		                       &$(0.6-2)\times 10^5$\\
S                                             & $(0.6-3)\times10^{8}$  &$ 5 \times 10^{13}$	&$ 10^{14}$ 		&$4 \times (10^{13}-10^{14})$           & $2 \times (10-10^{3})$\\
$d_i/{\textbf L}$                      & $5 \times10^{-9}$ 	   &$4 \times10^{-7}$		& $10^{-6}	$               &$7\times (10^{-3}-10^{-2})$                &  $0.1-1$     \\
$d_e$/{\textbf L}                     &  $10^{-10}$ 	           &$8 \times 10^{-9}$			&$3 \times 10^{-8}$	  &$2 \times (10^{-4}-10^{-3})$           &$(0.3-2) \times 10^{-2}$\\
\textbf{$a/{\textbf L}$}             &$(2-3)\times10^{-3}$   &$4 \times10^{-5} $	      &$5 \times 10^{-5}$     &\textcolor{red}{(0.4-1) $\times 10^{-3}$}  &\textcolor{red}{0.2-0.5}\\ 
\bf{$\delta_{\eta}/{\textbf L}$}  & $ (0.6-1) 10^{-4}$	   &$10^{-7}$                    	&$ 10^{-7}$	          &$(0.5-2) \times10^{-7}$               &$ (0.3-2) \times 10^{-1}$  \\
\hline\hline
\end{tabular}
\label{tabella1}
}
\caption{Characteristic plasma parameters of magnetized plasma environments, where Hall reconnection may occur.  Physical quantities are expressed in cgs units. Bold quantities are measured (or empirically derived), while S is the Spitzer resistivity. We assume {\bf T}$=T_e \sim T_i$ and {\bf n} $= n_e\sim n_i$. The inner resistive layer is estimated as $\delta/{\textbf L} \sim S^{-1/2}$, while the inverse aspect ratio is estimated as $a/{\textbf L} \sim S^{-1/3}$, where $d_i/\delta=P_h<1$, and as $a/{\textbf L} \sim S^{-1/3} P_h^{0.29}$, where  is $d_i/\delta=P_h\ge1$. For values in the solar corona as well as for the chromospere we refer to \citet{Aschwanden}, {\bf L} being the length scale of a flux tube and the thickness of the chromosphere respectively. For fast solar wind parameter (at 1 A.U.) we refer to \citet{EsserHabbal:1995}, considering the case where only protons and $\alpha$ particles are heated, and $L$ is the mean free path. For magnetotail reconnection parameters \citet{Ang,Sergeev,Kivelson}, typical conditions in the plasma sheet during a substorm growth phase have been considered. For MRX experiment see \citet{Yamadaetal:2014}. Red color means $a \le d_i$ which means the growth rate should be normalized with the whistler time.}
\end{table*}
We can see that for particular ranges of plasma parameters in the solar corona or solar wind for instance, the condition $\delta/L \sim h$ is verified; in Eq.\eqref{aspectratio}, using Eq.\eqref{exponentshall}, the factor $P_h^{\zeta} > 1$ when $P_h > 1$, so the trigger to an ``ideal" tearing instability occurs at larger aspect ratios than the resistive case. 

\section{Comments and conclusions.}

\begin{figure}
\centering
\includegraphics[width=83mm]{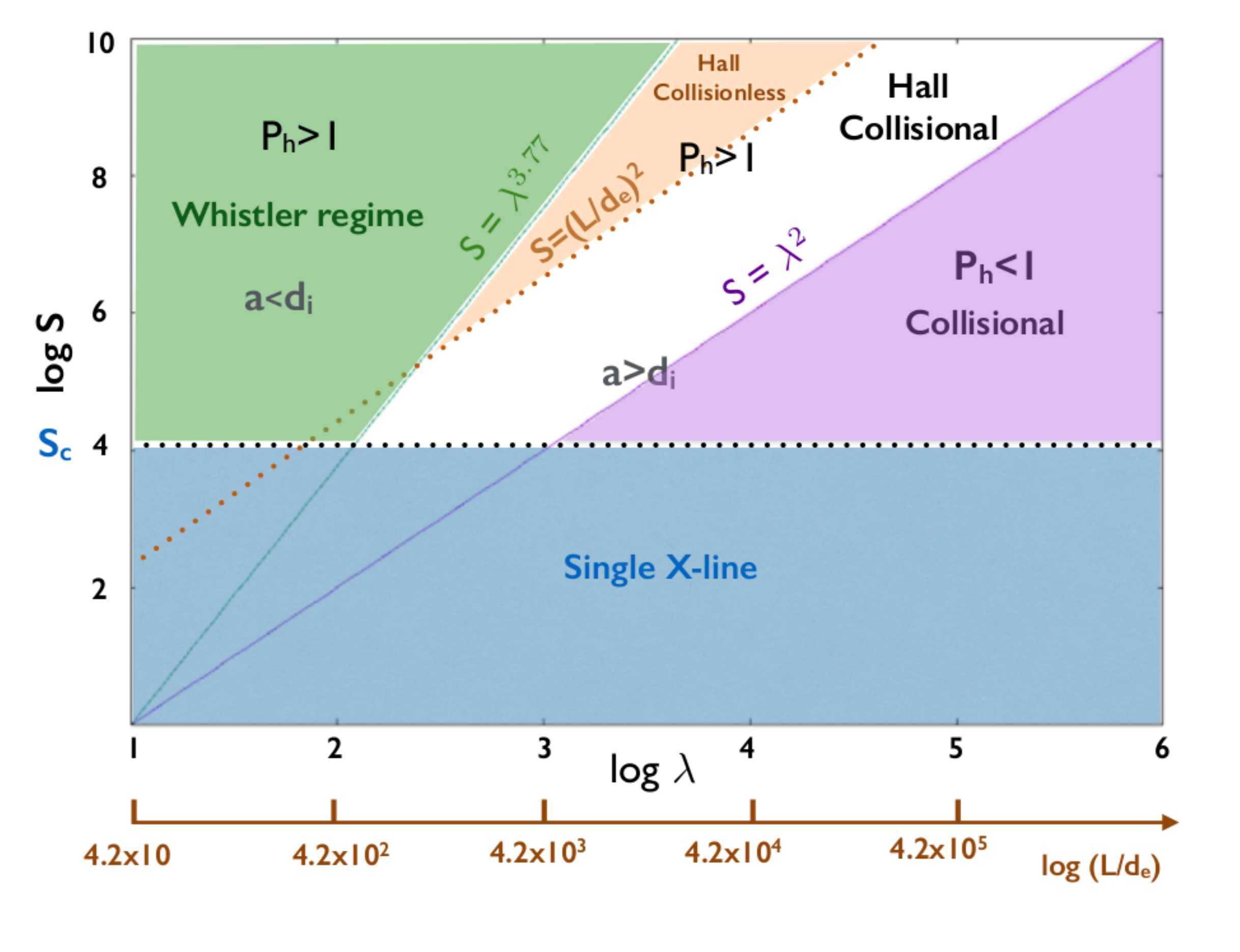}
\caption{Revision of the phase diagram in \citet{Ji:2011} on the base of the results in \citet{PucciVelli2014}, \citet{DelSartoetal2016} and the result from this paper.}
\label{phasediagram}
\end{figure}

We have studied the linear resistive tearing instability for current layers whose thickness approaches the ion inertial length, in which the  the instability growth rate and parameters are modified by the Hall effect. We have generalized the ``ideal" tearing criterion, taking into account a finite ion inertial length, ending up with a trigger relation for the aspect ratio which varies in the parameter space $(S , P_h)$, depending on the specific plasma parameters. The result is a couple of values $\alpha$ and $\beta$ defining a critical inverse aspect ratio, scaling as $a/L \sim S^{\alpha} P_h^{\beta}$, below which the reconnection process becomes explosive. Recently phase diagrams involving Lundquist number and the macrosopic system size in units of the ion inertial lenght (or ion sound gyroradius, if a guide field is present) have been created, which summarize the essential dynamics of the plasma for a wide range of parameters \citet{Ji:2011}. Fig. \ref{phasediagram} summarizes the results obtained using the IT criterion, in the absence of a guide field (in the figure,  $\lambda=1/h$). The $(S , \lambda)$ parameter space has an extension on the left due to the fact that we also take into account the effect of finite electron skin depth in the RMHD case discussed in \citet{DelSartoetal2016}, where $d_i=\sqrt{m_i/m_e}\, d_e \sim 42 d_e$. Indeed the Hall effect on its own can not break the frozen in conditions, i.e. collisionless reconnection must be triggered by other effects, and the inertial terms in Ohm's law are proportional to $d_e$. As in \citet{Ji:2011} we have a region (blu one) where the single X-line reconnection occurs  i.e. for Lundquist numbers smaller than the critical one determined by inflows and outflows, see e.g. \citet{TeneraniPosterAGU2016, ChenPosterAGU2016}. We have investigated the purple region in \citet{PucciVelli2014} and the white region within this work. While the orange region has been investigated in \citet{DelSartoetal2016} (even if the Hall effect is negligible in the adopted frame), the green one still has to be explored and with that, a possible critical value of $d_e$ for which collisionless reconnection can present single or multiple x-points. 
These are useful to understand the ongoing process regime in astrophysical plasmas as well as in laboratory plasmas, and in the latter case to visualize the parameter space of an experimental facility, and in particular can be applied to the multi-point observations of MMS, to compare theoretical predictions the spatial structure of the Hall magnetic and electric fields surrounding the diffusion region.
The next step for linear studies is to include the 3D structures which naturally arise in the presence of a mean magnetic field in the direction orthogonal to the plane where magnetic reconnection occurs. In this case other effects may occur due the dependence on the third direction and the system of equations would be more complicated, involving also the complex component of the eigenfunctions.
Then, electron pressure terms should be included. Such terms will introduce the dependence of critical aspect ratio also on the thermal ion gyroradius. From the non-linear evolution point of view, it would be of great interest to follow a collapsing current sheet (initially with 0 guide field), into the Hall regime. The resistive internal singular layer, which in 2 Dimensions becomes the thickness of secondary current sheets in nonlinear evolution (\citet{Teneranietal2016}), and where the Hall magnetic field is different from zero, tends to shrink for larger values of the Hall parameter, making it a challenging problem which we hope to address in the near future.

\section{Acknowledgements}
This work was supported by the NASA Solar Probe Plus Observatory Scientist grant and the NSF - DOE partnership in basic plasma science and engineering.

\bibliography{biblyography}

\end{document}